\documentclass[aps,pre,twocolumn,superscriptaddress]{revtex4-1}

\usepackage{bm}
\usepackage{epsfig}
\usepackage{amsmath}
\usepackage{graphicx}
\usepackage{amssymb}
\usepackage[normalem]{ulem}
\usepackage[usenames]{color}
\begin{document}

\newcommand{\kbt}{k_\mathrm{B} T}
\newcommand{\kb}{k_\mathrm{B}}
\newcommand{\kt}{k_\mathrm{B}T}
\newcommand{\rhoinf}{\rho_\infty}
\newcommand{\rhoc}{\rho_\text{c}}
\newcommand{\rhoAinf}{\rho_{\text{A}\infty}}
\newcommand{\rate}{\mathcal{R}}
\def\Rs{R_\text{s}}

\title{Diffusion-limited rates on low-dimensional manifolds with extreme aspect ratios}
\author{Aleksandr Kivenson}
\affiliation{Department of Biochemistry, Brandeis University, Waltham, MA}
\author{Michael F. Hagan}
\email{hagan@brandeis.edu}
\affiliation{Department of Physics, Brandeis University, Waltham, MA}
\date{\today}

\begin{abstract}
We consider a single-species diffusion-limited annihilation reaction $2A\rightarrow \varnothing$ with reactants confined to a two-dimensional surface with one arbitrarily large dimension and the other comparable in size to interparticle distances. This situation could describe reactants which undergo both longitudinal and transverse diffusion on long filamentous molecules (such as microtubules), or molecules that undergo truly one-dimensional translational diffusion (e.g. a transcription factor on DNA) but simultaneously exhibit diffusive behavior in a second dimension corresponding to a rotational or conformational degree of freedom. We combine simple analytical arguments and Monte Carlo simulations to show that the reaction rate law exhibits a crossover from one-dimensional to two-dimensional diffusion as a function of particle concentration and the size of the smaller dimension.  In the case of a reversible binding reaction, the diffusion limited reaction rate is given by the Smoluchowski expression, but the crossover is revealed in the statistics of particle collision histories. The results can also be applied to a particle-antiparticle annihilation reaction $A+B\rightarrow\varnothing$.

\end{abstract}

\maketitle

\section{Introduction}
\label{sec:intro}
This paper studies a diffusion-limited single-species annihilation reaction \mbox{$2A\rightarrow\varnothing$} on a finite two-dimensional manifold for which one dimension is macroscopic in size and the size of the second is comparable to interparticle distances. Collisions between particles confined to this manifold lead to the formation of reaction products which are either inert or irreversibly dissociate from the manifold. Diffusion-limited  annihilation reactions  play important roles in a variety of physical and chemical systems (e.g.\ \cite{avnir84}) and the kinetics of particle annihilation in low dimensions has been studied extensively in the context of particle-antiparticle annihilation (e.g.\ \cite{krapivsky95,toussaint83}). It has been shown that correlations among particle positions render the traditional Smoluchowski expression for the diffusion-limited rate incorrect for systems with fewer than three dimensions, and different rate laws have been established for one or two macroscopic dimensions. However, the effect of boundaries on low-dimensional reaction rates has not been thoroughly investigated, and it remains unclear what rate law applies for systems with extreme aspect ratios.

While our results can be applied to particle-antiparticle annihilation, we were specifically motivated to consider reactions on manifolds with disparate dimensions because many reactions in biological systems take place on one- or two- dimensional scaffolds, such as lipid-bilayer membranes, microtubules, or nucleic acids.  It is becoming apparent that biological systems exploit low-dimensional manifolds to increase the rate at which molecules collide beyond the diffusion limited rate for molecules in three dimensions (3D). For example, extensive theoretical work (e.g. \cite{Berg1981,Hu2006,Richter1974}) and \emph{in vitro} experiments (e.g. \cite{Gowers2005,Jeltsch1996,Ruusala1992,Blainey2006,Bonnet2008} and reviewed recently in \cite{gorman08}), and even \emph{in vivo} measurements \cite{Elf2007,Hammar2012} on transcription factors show that augmenting 3D diffusion with one-dimensional (1D) sliding of transcription factors on DNA molecules enhances the rate at which they find specific target regions on DNA. Similar increases in reaction rates have been identified for  membrane-bound G-protein signaling (reviewed in \cite{baker07}),  and it has been proposed that the sliding of viral capsid protein molecules along an RNA molecule enhances the rate of capsid assembly \cite{elrad10, Hu2007b, Kivenson2010}.

We thus consider a physical system in which particles diffuse on a filamentous molecule such as a long microtubule or nucleic acid, and undergo bimolecular reactions to form complexes that either dissociate from the filament or become otherwise inert. Our analysis can also describe molecules that diffuse and react with specific sites on the filament, such as a transcription factor searching for a specific region of sequence. In the systems under consideration, the macroscopic dimension corresponds to the longitudinal direction along the filament. The short dimension can correspond to the lateral direction if molecules undergo transverse diffusion (e.g. on a microtubule). Alternatively, for molecules that undergo true one-dimensional translational diffusion (e.g. along a DNA molecule), the short dimension can correspond to a rotational or conformational degree of freedom whose value determines the reaction probability. For example, cross-sections for protein association reactions depend critically on the relative orientations of the colliding reactants, and recent measurements indicate that a transcription factor has a relatively small probability of binding to its operator region each time it slides over it \cite{Hammar2012}. We show that the reaction rate law in such a system exhibits a density-dependent crossover from one-dimensional to two-dimensional behavior as the size of the bounded dimension increases.

The important implication of this result is that experimentally measured rate constants in such a system will  depend on particle concentration for some system parameters, with a form that itself varies with concentration. Our prediction of a crossover between one- and two-dimensional behavior also applies to particle-antiparticle annihilation reactions. Furthermore we show that for reversible diffusion-limited reactions, the particle correlations leading to alternative rate laws in low dimensions for irreversible reactions are reflected in the histories of particle-pair identities.

\section{Theory}
\subsection{Unbounded dimensions}
We consider a system of $N$ particles which have diffusion constant $D$ and annihilate upon collision; i.e. $2A\rightarrow \varnothing$.  For the moment consider a system in $d$ dimensions, each of which has macroscopic size $L$.

Traditionally the diffusion-limited rate $\rate$ is calculated under the ``mean field'' approximation that particle concentration \mbox{$\rho=N/L^d$} is uniform in space and thus the bimolecular collision rate is given by $\rate=d\rho/dt = k \rho^2/2$. The rate constant \mbox{$k=16\pi R D$} with $R$ the particle radius can be obtained from the  Smoluchowski equation as described in the appendix. However, for fewer than three dimensions  the inefficiency of the diffusion mechanism leads to amplifications of initial local density fluctuations and the reaction rate takes different,  density-dependent forms. This was shown using dimensional analysis by Toussaint and Wilczec \cite{toussaint83} and Krapivsky \cite{krapivsky95} as follows. In the mean field approximation it can be assumed that the rate constant depends on the particle radius and the diffusion constant, $k=k(D,R)$, in which case dimensional analysis leads to $k \propto DR^{d-2}$ with $d$ the number of dimensions. To be physically realistic, the rate constant must increase with radius $R$, leading to the conclusion that the mean field expression is only valid for $d>2$. For $d\le2$ we must assume that the rate constant depends on the particle density rather than the radius, $k=k(D,\rho)$, in which case dimensional analysis leads to $k\propto D \rho^{-1+2/d}$ and thus the reaction rate is given by $\rate \propto D \rho^{1+2/d}$ for $d\le2$. Since $d=2$ is the marginal dimension, logarithmic corrections to the mean field expression arise \cite{krapivsky95}, which we derive approximately in the appendix. The dimension dependent reaction rates can then be summarized as
\begin{align}
\rate_\text{1D} & = A_1 D \rho^3 \label{eq:rate1D}  \\
\rate_\text{2D} & = \frac{ A_2 D \rho^2}{\ln\left(B/\rho\right)} \label{eq:rate2D} \\
\rate_\text{3D} & = 8 \pi R D \rho^2
   \label{eq:rate3D}
\end{align}
 where $A_1$, $A_2$, and $B$ are unknown scaling constants, with $A_1$ and $A_2$ determined by $D$ and the collision cross-sections.

\subsection{Bounded dimension}
The expressions for the diffusion limited rates (Eqs. \ref{eq:rate1D}-\ref{eq:rate3D})  apply to a manifold that is unbounded in each dimension. We now address the question of what happens on a manifold in which one of the dimensions is finite.
We consider $N$ particles in a 2D system with dimensions of size $L$ and $w$ with $L>>w$. Motivated by the physical systems described in the introduction, we consider the short dimension to be periodic, but the results are qualitatively unchanged for reflecting boundary conditions. Without loss of generality we set the diffusion constants in each dimension equal to $D$; the size of the second dimension can always be rescaled to achieve this situation.

For sufficiently large $w$ the particles are unaffected by the boundary conditions and we expect eq.\,\eqref{eq:rate2D} to apply, with $\rho=N/Lw$. However,  below a threshold value of $w$, particles cover space in the short direction more rapidly than they collide. In this limit the short dimension size will play no role, and the  reaction rate will follow $r=A_1 D \rho_\text{1D}^3$  with  the effective one-dimensional concentration $\rho_\text{1D}=N/L$.  We expect a crossover between the one- and two-dimensional rate laws at the threshold value of $w$. A mean field argument would suggest that the crossover occurs at the particle concentration for which the typical distance between particles is equal to the short dimension size, $\rho_\text{1D}=w^{-1}$; however, the correlations between particle positions described above change this estimate dramatically. Instead, noting that the frequency of collisions can never exceed that predicted by either equation \eqref{eq:rate1D} or \eqref{eq:rate2D}, the crossover can be determined as the short dimension size $w$ and 2D particle concentration $\rhoc$ for which collision frequencies predicted by the  1D and 2D expressions are equal:
\begin{align}
\rhoc \ln\left(\frac{B}{\rhoc}\right) = \frac{A_2}{A_1}\frac{1}{w^2}
\label{eq:crossover}.
\end{align}
This transcendental equation can be solved numerically for $\rho$ given $w$, $L$, with the parameters $B$ and $A_2/A_1$ determined independently from measured reaction rates.

\subsection{Reversible diffusion-limited reactions}
\label{sec:reversibleTheory}
It is important to note that the equations described above apply to nonequilibrium irreversible reactions. In the case of a reversible diffusion limited reaction, where particles form dimer complexes with finite lifetimes, the mean-field prediction that reaction rate will be proportional to the square of the concentration must be correct, since the particle positions are distributed uniformly in the absence of energetic interactions to satisfy the Boltzmann distribution \cite{Sinder2000,*Koza2003a,*Chopard1993}.
However, the nature of diffusion in low dimensions and the resulting correlations described above are reflected in the history of particle collisions. Specifically, the rate of ``novel collisions'', meaning collisions between pairs of particles which have never previously interacted, will follow the forms given in \eqref{eq:rate1D} or \eqref{eq:rate2D}. With particle tracking techniques and or single molecule fluorescence experiments \cite{Bonnet2008,Blainey2006,Elf2007,Hammar2012} it is possible to monitor the history of particle collisions in a reversible system. We show below that the expressions \eqref{eq:rate1D} and \eqref{eq:rate2D} are nearly recovered even if particle collisions are monitored only to first order, i.e., a collision between particles $i$ and $j$ is defined as ``first-order novel'' so long as particle $j$ is not the last particle that $i$ was bound to and vice versa.

\section{Simulations}

To investigate the predicted rate laws \eqref{eq:rate1D} and \eqref{eq:rate2D} and the density-dependent crossover size \eqref{eq:crossover}, we performed Monte Carlo simulations to study a diffusion-limited annihilation reaction on a two-dimensional rectangular lattice with periodic boundary conditions, with side lengths $L$ and $w$ measured in units of the lattice spacing $a$. In each Monte Carlo move a particle was chosen randomly and moved to one of its four neighboring lattice sites, chosen at random. If the move placed the chosen particle on an already occupied lattice site, the two particles annihilated each other (i.e. they were deleted from the system). We measure distances in units of $a$ and time in units of Monte Carlo sweeps, defined as the number of moves required  for each particle to make on average one move in the long direction. Thus the diffusion constant was $D=1/2$. 

Since the model was intended to investigate the effects of correlations and particle positions in a nonequilibrium system, a statistically appropriate initialization of particle positions was essential. We began all simulations with 100\% of lattice sites occupied by particles, in which case there was only one possible  arrangement and thus no \emph{a priori} assumptions about the distribution were necessary. We then measured the annihilation rate as a function of the number of particles $N(t)$ until the particle number decreased to near zero.

We considered a second set of simulations to test the assertions corresponding to reversible collision-limited reaction rates. These simulations were performed as described for the annihilation simulations, except that when a Monte Carlo move would transfer a particle on to an already occupied lattice site, the initial positions of the two particles were swapped. This simulated formation of a complex whose lifetime is short in comparison to the characteristic timescale for a particle to diffuse its diameter. The results pertaining to the form of the reaction rate law did not change when longer complex lifetimes were considered. In these simulations the previous reaction partner of each particle was stored in memory, and we recorded both the overall rate of collisions and the rate of first order novel collisions as defined above. These simulations were performed for a range of particle numbers $N$.

In most simulations we set the long dimension size $L=10^6$, but to enable sufficient statistics at low densities we performed additional simulations with $L=10^7$. We varied $w$ in the range $w\in[5,500]$. We performed 10 -- 100 simulations for each value of $w$. The simulations were optimized to efficiently handle large systems with concentrations ranging from 100\% to near 0\%.

\section{Results}

\begin{figure}
\centering
\includegraphics[width=0.99\columnwidth]{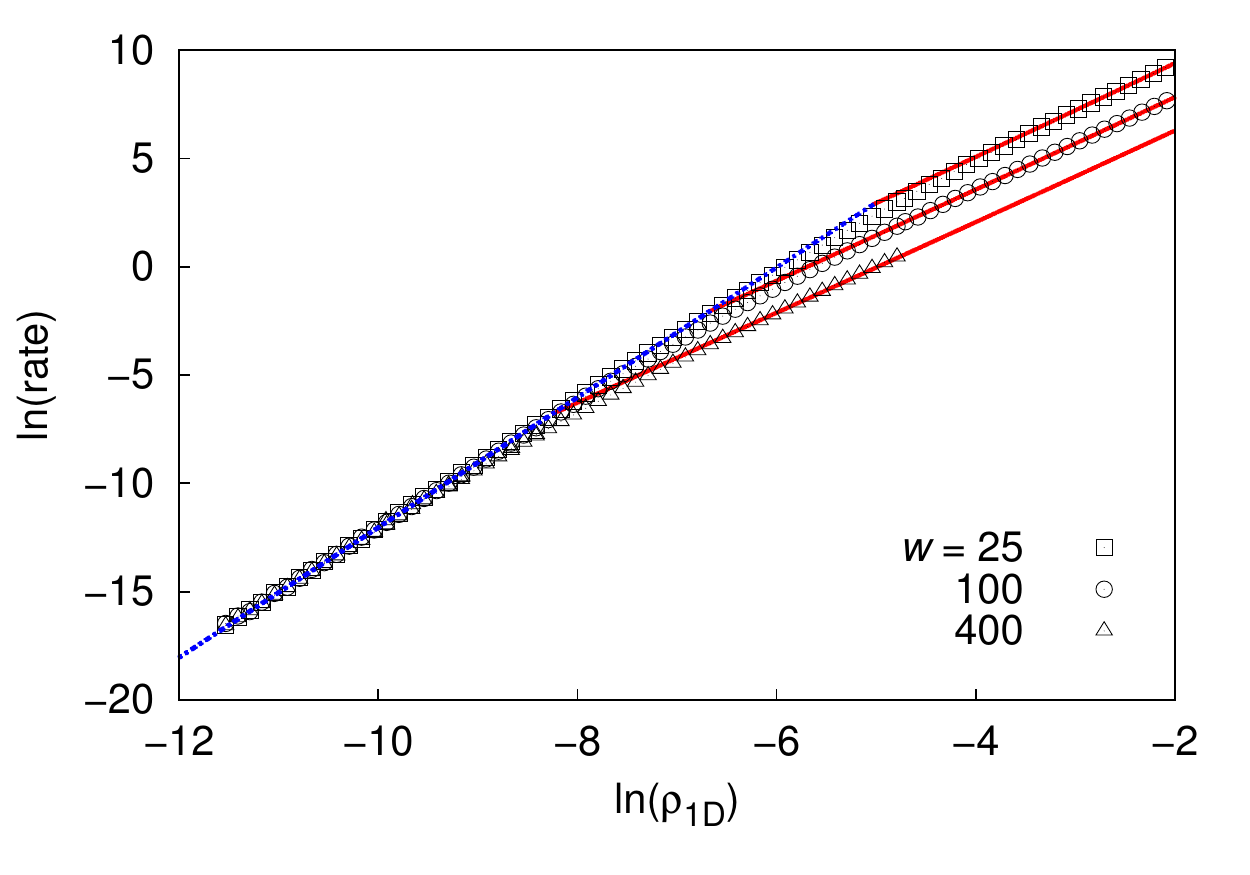}
\caption{\textbf{Simulations for all short dimension sizes $w$ demonstrate a crossover from 2D to 1D behavior, and that annihilation rates in the 1D regime do not depend on $w$.}  Color online. Annihilation rates as a function of the effective one-dimensional concentration $\rho_\text{1D}=N/L$ are shown for three representative values of $w$.  The red solid lines and blue dashed line respectively show fits to the 2D (Eq. \eqref{eq:rate2D}) and 1D (Eq. \eqref{eq:rate1D}) rate laws.  }
\label{fig:exampleCrossover}
\end{figure}

\begin{figure}
\centering
\includegraphics[width=0.99\columnwidth]{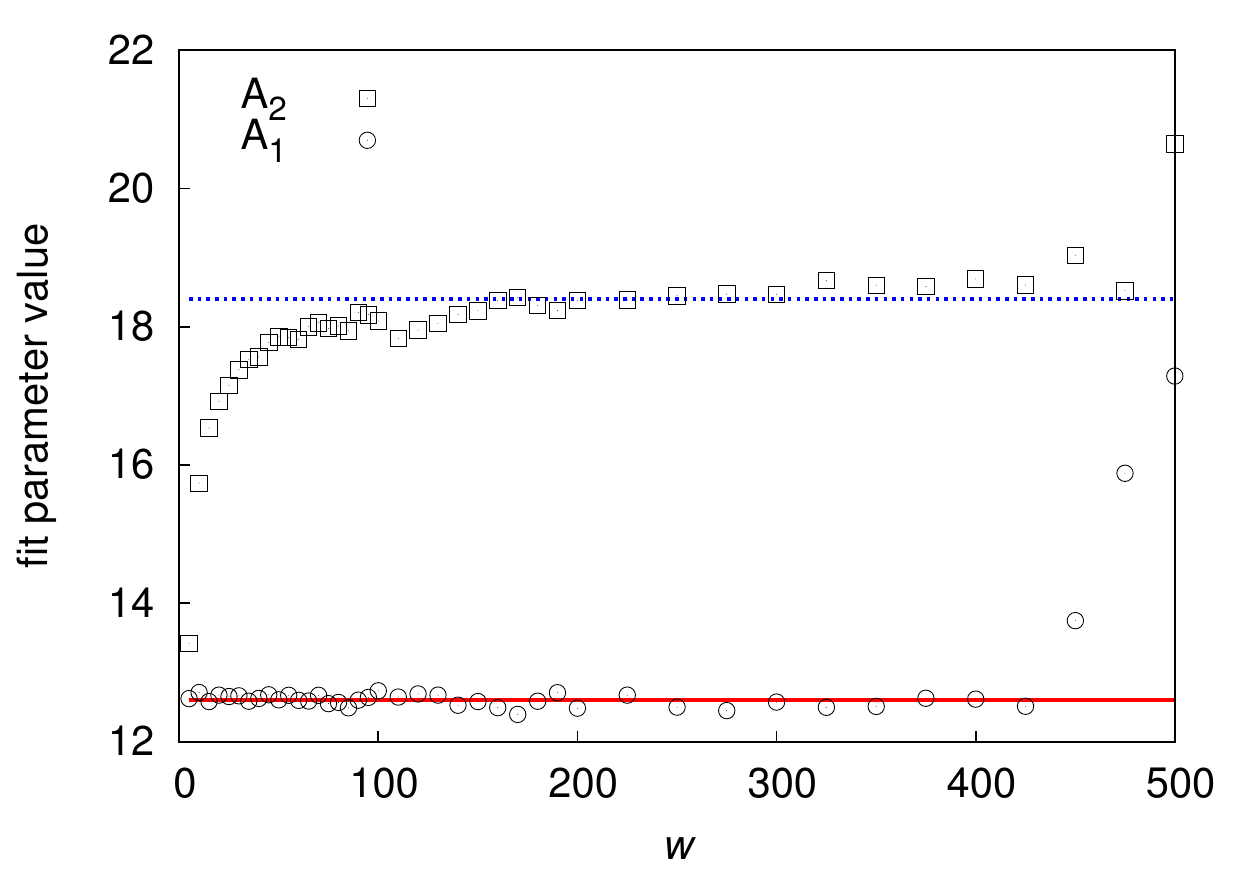}
\caption{\textbf{Fit parameter values are independent of $w$.} Color online. The fitted values for $A_1$ and $A_2$ are shown as functions of short-dimension width $w$. They are approximately constant over the simulated range of $w$, except at the extremes where statistics are limited.  The constant values for $A_1$ and $A_2$ used for the global fit in Fig. \ref{fig:allCrossovers} are shown respectively as a solid red line and a dotted blue line.}
\label{fig:diffConsts}
\end{figure}

Simulations for all sizes $w$ of the short dimension demonstrate a crossover from the 2D diffusion-limited rate law at high concentrations (i.e.\ large values of $N$) to 1D scaling at low concentrations. Furthermore, the annihilation rates at different values of $w$ below the crossover concentration collapse onto a single curve, consistent with the expectation of one-dimensional behavior.  To illustrate these observations, annihilation rates as a function of the effective one-dimensional concentration $\rho_\text{1D}$ are shown for three short dimension sizes in Fig.~\ref{fig:exampleCrossover}. We see that  the measured reaction rates at small or large densities agree very well with the theoretical rate laws for one and two dimensions, \eqref{eq:rate1D} and \eqref{eq:rate2D} (shown as solid red and dashed blue lines, respectively).  The data was fit to the theoretical expressions independently at each value of $w$, with fit parameters $A_1$ for $\rate_\text{1D}$ and $A_2$ for $\rate_\text{2D}$. The parameter $B$ for $\rate_\text{2D}$ is also unknown, but we held it constant at $B=1$ to reduce the number of fit parameters. Further evidence that the theories match the simulation data is shown in Fig. \ref{fig:diffConsts}, where we see that the independently fit values of $A_1$ and $A_2$ are constant within error over a wide range of $w$.

\begin{figure}
\centering
\includegraphics[width=0.99\columnwidth]{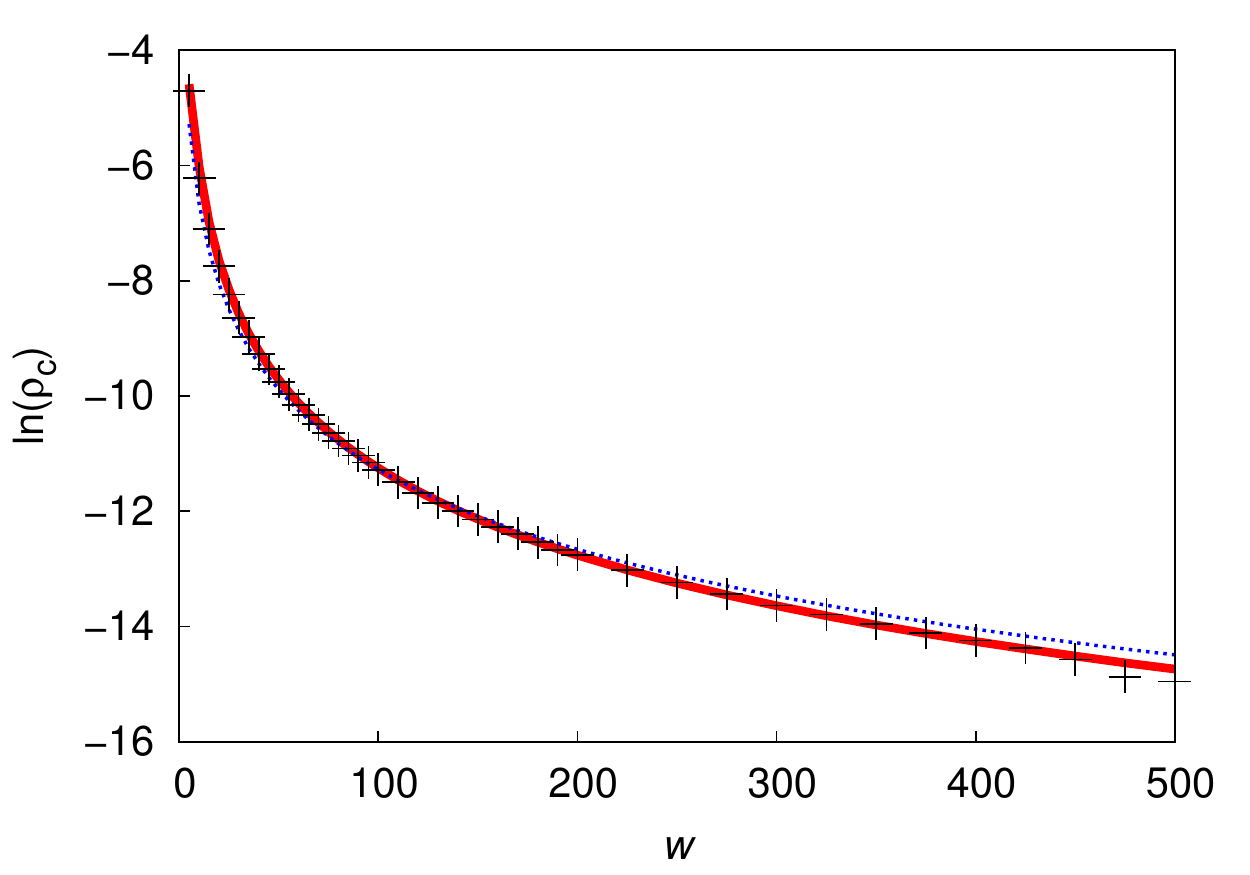}
\caption{\textbf{Measured crossover densities match theoretical predictions.} Color online. The measured crossover concentration $\rho_\text{c}$ corresponds well to Eq.~\eqref{eq:crossover} solved using constant values of $A_1 = 12.6$ and $A_2 = 18.4$ (solid red line, see Fig. \ref{fig:diffConsts}) over the entire range of short-dimension sizes $w$. The simulated crossover densities ($+$ symbols) are obtained from averages over many independent simulations at each value of $w$ as described in the text. If the logarithmic correction to 2D behavior is neglected, the fit worsens (the best attempt is shown as a blue dotted line).}
\label{fig:allCrossovers}
\end{figure}

Near the crossover, the measured rate is smaller than that predicted by either expression, as assumed in the theoretical analysis.  The fact that the 1D and 2D theoretical rate laws fit the data for small and large $w$ respectively, with essentially constant values of $A_1$ and $A_2$, then ensures that the theoretical prediction for the crossover density, $\rhoc$ in Eq.~\eqref{eq:crossover}, will provide a good match for the data. To illustrate this, we define the simulated crossover densities as the densities for each value of $w$ at which the rate laws fit respectively to small and large $N$ intersect. In Fig.~\ref{fig:allCrossovers} we then compare these simulated crossover densities to the the theoretical prediction Eq.~\eqref{eq:crossover} calculated using constant values of $A_1=12.6$ and $A_2=18.4$.
 We see that the correspondence between the simulated and theoretical crossover densities is excellent over two decades of short dimension sizes.  We also show in the figure the best fit for the crossover densities if the logarithmic correction factor for the two-dimensional regime is not included in Eq.~\eqref{eq:rate2D}. This expression is slightly but consistently less accurate.

\subsection{Reversible reaction rates}
\begin{figure}
\centering
\includegraphics[width=0.99\columnwidth]{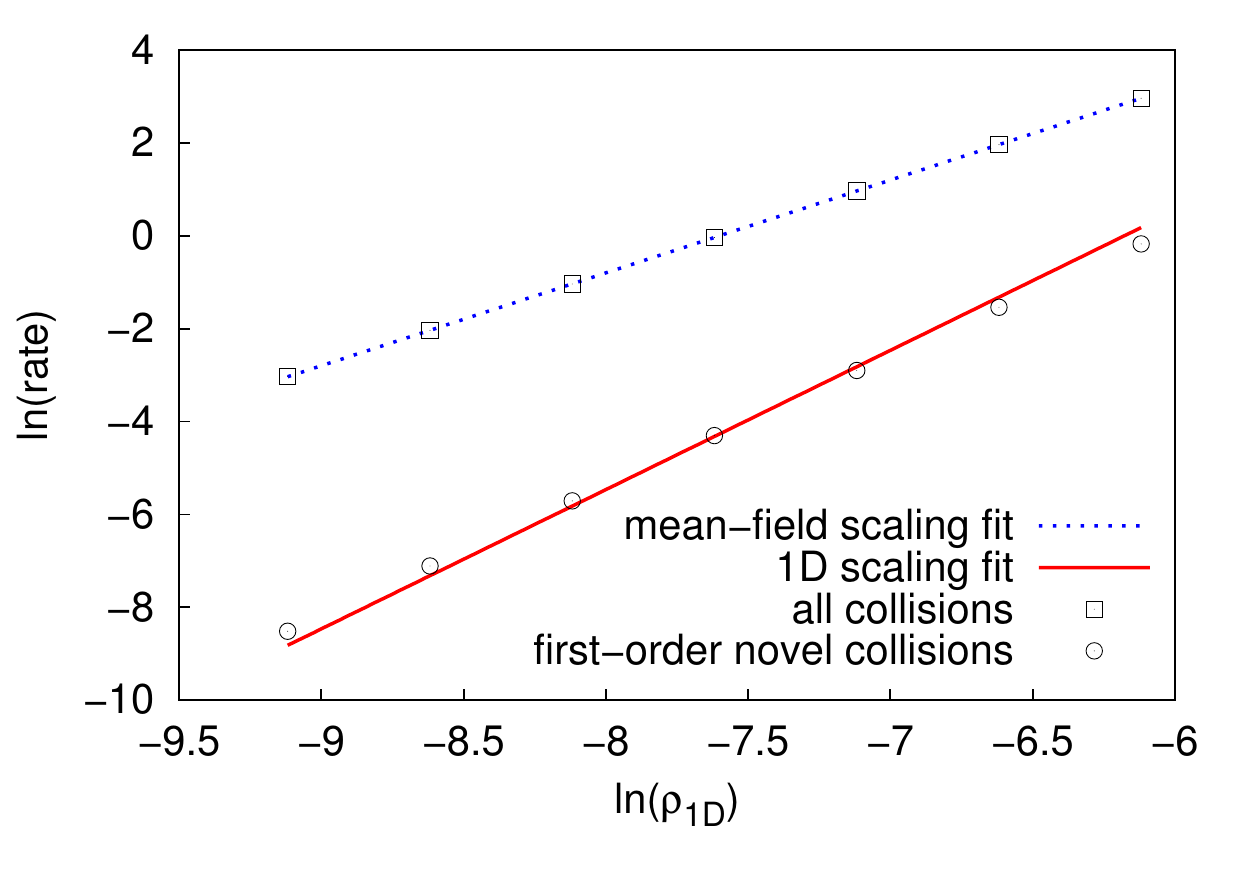}
\caption{\textbf{Reversible reaction rates follow mean-field scaling, but first-order novel collision rates approach the irreversible scaling.} Color online. Collision rates are shown for a system in which particles swap places rather than annihilate upon colliding, with \(w = 10\) over a range of effective one-dimensional concentrations $\rho_\text{1D}$ well below the crossover density.  The overall collision rate (red solid line) obeys the mean-field approximation for two dimensions ($\boxdot$).  However, the rate of ``first-order novel'' collisions ($\odot$), defined in the text, nearly restores the 1D irreversible collision rate scaling (dashed blue line).}
\label{fig:mfDiffusion}
\end{figure}

Reaction rates for the reversible model in which particles swap rather than annihilate upon colliding were measured at $w=10$ over a range of densities well below the crossover density $\rhoc=2.0\times10^{-3}$. In contrast to the irreversible annihilation model, which behaved according to the one-dimensional diffusion limited rate law at these densities, the overall reaction rates for this reversible system are well fit by the mean field rate expression $\rate\cong D \rho^2$ (Fig.~\ref{fig:mfDiffusion}). Interestingly though,  the one dimensional law is nearly recovered when the rate of first-order novel collisions (defined in section~\ref{sec:reversibleTheory}) is plotted as a function of density.

\section{Conclusions and outlook}
The results of our simulations demonstrate that diffusion-annihilation reactions on two-dimensional manifolds with large aspect ratios transition from one-dimensional behavior to two-dimensional behavior as the size of the shorter dimension crosses a threshold value. The simulation results are well fit by our theoretical expressions, and confirm the significance of the logarithmic correction that arises in the diffusion-annihilation rate law in two-dimensional systems. Furthermore, our simulations of the reversible reaction system show that particle correlations and the resulting transition to one-dimensional behavior at extreme aspect ratios are reflected in the history of reaction partners.

While we consider one macroscopic dimension and one small dimension, extension of our results to other situations is straightforward. For example, in the case of one macroscopic dimension with length $L$ and two small dimensions with sizes $w$ and $h$ (e.g. a particle diffusing within a narrow tube, or a particle undergoing translational diffusion on a filament and two rotational dimensions) and a particle density $\rho=N/Lwh$, we expect a crossover from one-dimensional to three-dimensional behavior at density $\rho \cong R /d^2 w h$. Similarly for diffusion and reaction of particles on a membrane, a single rotational dimension leads to a crossover between two-dimensional and three-dimensional (i.e.\ mean field) behavior. The important observation is that the crossover size of the ``small'' dimensions can become quite large for sufficiently small particle densities.

Finally, extension of our results to particle-antiparticle annihilation, $A+B\rightarrow \varnothing$, is straightforward. While this reaction and the single species annihilation we consider here yield different kinetics for inhomogeneous starting conditions \cite{krapivsky95}, the crossover relationship we predict is unchanged except within scaling constants.

\section{Acknowledgements} This work was supported by the NIH through Award Number R01AI080791 from the National Institute Of Allergy And Infectious Diseases; MFH was also supported by NSF-MRSEC-0820492. We thank Pavel Sountsov and Jeff Bombardier for performing initial simulations that motivated this work.  We gratefully acknowledge computational support provided by the Brandeis HPC and the NSF XSEDE facilities (Purdue Condor).

\appendix
\section{Diffusion limited rates in one and two dimensions}

In this section we obtain scaling laws for the diffusion limited rate in one and two dimensions by two approaches.

\subsection{Fick's Law approach}

First, we follow the classical solution for the diffusion-limited rate in three dimensions. In this approach, the rate constant for bimolecular collisions of particles each with radius $R$ and diffusion constant $D$ is mapped on to the problem of calculating the flux of point particles with diffusion constant $D'=2D$ to an infinitely absorbing sphere with radius $R_\text{s}=2R$ \cite{berg77}. The concentration of point particles is held fixed at $\rhoinf$ at a radius $b>R$. We then seek the steady-state flux of point particles to the sphere by solving the steady-state solution of the diffusion equation with $\rho(\mathbf{r})$ the concentration of point particles at position $\mathbf{r}$:
\begin{align}
\nabla^2 \rho(\mathbf{r})=&0 
\label{eq:laplace}
\end{align}
with $\rho(R_\text{s})=0$, $\rho(b)=\rhoinf$ and $r=|\mathbf{r}|$.  In three dimensions with spherical symmetry this yields
\begin{align}
\rho(r)=\rhoinf\frac{1-\Rs/r}{1-\Rs/b}
\label{eq:conc}.
\end{align}
The reaction rate constant $k_\text{3D}$ is then given by the area of the absorber, $4\pi \Rs^2$, times the inward flux, $-J(R)$, to give
\begin{align}
k_\text{3D} = 4\pi \Rs \rhoinf D \frac{1}{1-\Rs/b}
\label{eq:rateone}.
\end{align}
The usual expression for the diffusion limited rate is obtained by setting $b=\infty$, substituting $D'=2D$ and $\Rs=2R$ and assuming the mean-field expression $\rate_\text{3D}=k_\text{3D} \rhoinf/2=8\pi R D \rho^2$.

Following the same approach in one or two dimensions yields (assuming $b\gg R$):
\begin{align}
k_\text{2D} =& \frac{4\pi D \rhoinf}{\ln(b/2 R)} \nonumber \\
k_\text{1D} =& 2D \rhoinf / b
\label{eq:ratelowda}
\end{align}

One immediately sees that the classical approach for three dimensions breaks down because there is no steady-state solution in the limit $b\rightarrow\infty$. The situation can be rescued, however, by realizing that within the mean field assumption the bulk concentration must be established within a distance $b\cong\rhoinf^{-1/d}$ from any molecule with $d$ the number of dimensions. Substituting this relationship into  \eqref{eq:ratelowda} yields
\begin{align}
k_\text{2D} =& \frac{8\pi D \rhoinf}{\ln(1/4 \rhoinf R^2)} \nonumber \\
k_\text{1D} =& 2 D \rhoinf^2
\label{eq:rateconstantlowdb}
\end{align}
and thus
\begin{align}
\rate_\text{2D} =& 4\pi D \rhoinf^2\frac{1}{\ln(1/4 \rhoinf R^2)} \label{eq:ratelowd2D} \\
\rate_\text{1D} =&  D \rhoinf^3
\label{eq:ratelowd1D}
\end{align}
which are nearly equivalent to the expressions obtained by the more sophisticated quasi-static analysis that follows.

\subsection{Quasi-static approximation}

In this approach we begin by considering the rate expression for the diffusion-annihilation reaction $2A\rightarrow \varnothing$, with the density of species at position $\mathbf{r}$ and time $t$ given by $\rho(\mathbf{r,t})$ and mean density $\rhoinf(t)$, particle size $R$ and diffusion constant $D$. Following Ref. \cite{krapivsky95} we start with the fundamental expression for the reaction rate as $r\sim \rho/T$ with $T$ the typical time interval before a given particle experiences a collision. Now we consider a tagged particle  and work in a coordinate system with origin at the center of the tagged particle. Neglecting correlations among nearby particles with each other, their density  satisfies the time-dependent diffusion equation
\begin{align}
\frac{\partial \rho(\mathbf{r},t+\tau)}{\partial \tau} = 2 D \nabla^2_\mathbf{r}
\rho(\mathbf{r},t+\tau)
\label{eq:qsone}
\end{align}
with the factor of 2 because the tagged particle diffuses as well.

In the quasi-static approximation \cite{redner90, krapivsky95} it is assumed that density correlations over time $\tau$ with respect to the tagged particle are limited to the distance it sweeps out in space by diffusion, $\sqrt{2 D\tau}$. Thus we solve \eqref{eq:qsone} with boundary conditions $\rho(r=2R)=0$ for $\tau>0$ and $\rho(r\ge\sqrt{2 D\tau},\tau)=\rhoinf$ to obtain
\begin{align}
\rho(\mathbf{r},t+\tau)\approx \rho(\mathbf{0},t) \frac{2\ln\left(r/2R\right)}{\ln\left(D\tau/2R^2\right)}
\label{eq:qstwo}.
\end{align}

The collision time $T$ is then calculated as the time at which the flux across the circle of radius $r=2R$ equals unity:
\begin{align}
8 \pi D R \int_0^T \frac{\partial \rho(r=2R,t+\tau)}{\partial r} d\tau = 1
\label{eq:qsthree}.
\end{align}
Inserting \eqref{eq:qstwo} into \eqref{eq:qsthree} gives to leading order $T\approx \ln(1/32\pi\rhoinf R^2)/8\pi D\rhoinf$. We then obtain the reaction rate
\begin{align}
\rate_\text{2D}=4\pi D \rho^2\frac{1}{\ln(1/32 \pi \rho R^2)}
\label{eq:qsfour}.
\end{align}
The same analysis for $d=1$ yields \eqref{eq:ratelowd1D}.


%

\end{document}